\documentclass[a4paper,11pt]{article}
\usepackage{a4wide}
\usepackage{amsfonts}
\usepackage{amssymb}
\usepackage{amsthm}
\usepackage{amsmath}

\newtheorem{theorem}{Theorem}[section]

\newtheorem{proposition}[theorem]{Proposition}

\begin{document}

\title{Some Late-time Asymptotics of General Scalar-Tensor Cosmologies}
\author{John D. Barrow$\ $ \\
%EndAName
DAMTP, Centre for Mathematical Sciences, \\
University of Cambridge,\\
Wilberforce Road, Cambridge CB3 0WA, UK. \and Douglas J. Shaw \\
%EndAName
Astronomy Unit, Queen Mary University, \\
Mile End Rd., London E1 4NS, UK. }
\date{}
\maketitle

\begin{abstract}
We study the asymptotic behaviour of isotropic and homogeneous universes in
general scalar-tensor gravity theories containing a $p=-\rho $ vacuum fluid
stress and other sub-dominant matter stresses. It is shown that in order for
there to be approach to a de Sitter spacetime at large 4-volumes the
coupling function, $\omega (\phi )$, which defines the scalar-tensor theory,
must diverge faster than $|\phi _{\infty }-\phi |^{-1+\epsilon }$ for all $%
\epsilon >0$ as $\phi \rightarrow \phi _{\infty }\neq 0$ for large values of
the time. Thus, for a given theory, specified by $\omega (\phi ),$ there
must exist some $\phi _{\infty }\in (0,\infty )$ such that $\omega
\rightarrow \infty $ and $\omega ^{\prime }/\omega ^{2+\epsilon }\rightarrow
0$ as $\phi \rightarrow \phi _{\infty }$ in order for cosmological solutions
of the theory to approach de Sitter expansion at late times. We also
classify the possible asymptotic time variations of the gravitation
`constant' $G(t)$ at late times in scalar-tensor theories. We show that
(unlike in general relativity) the problem of a profusion of
\textquotedblleft Boltzmann brains" at late cosmological times can be
avoided in scalar-tensor theories, including Brans-Dicke theory, in which $%
\phi \rightarrow \infty $ and $\omega \sim o(\phi ^{1/2})$ at asymptotically
late times.

\end{abstract}

\section{\protect\bigskip Introduction}

The cosmological approach to de Sitter space-time at large expansion times
has assumed a double importance in the study of the universe. It provides a
description of the evolution of the early universe after the prolonged
gravitational influence of a slowly-evolving inflation field, whose energy
density remains approximately constant for a significant interval of time.
It also provides a good description of the late-time expansion dynamics of
the universe in the presence of a dominant source of dark energy. We know
that an exact de Sitter solution arises when the effective equation of state
has the form $\rho +p=0$, where $p$ is the total isotropic pressure, and $%
\rho $ is the total matter density. This can arise because of the presence
of a single vacuum `fluid' with this equation of state, as first suggested
by Lema\^{\i}tre in 1933 \cite{lem}, and employed in the construction of the
steady state universe by Hoyle \cite{hoyle} and McCrea \cite{mcrea}, and in
the deduction of its stable asymptotic behaviour by Hoyle and Narlikar \cite%
{hoynar}. It could also result from the presence of an imperfect fluid which
possess an effective equation of state of this form, as is the situation
with bulk viscosity \cite{jb2}, or from the presence of higher-order
curvature terms in the gravitational Lagrangian beyond those sufficient to
generate general relativity \cite{starob}. The correspondence between the
situation in general relativity and such higher-order gravity theories can
be understood in terms of the conformal equivalence of the two theories \cite%
{conf}. The situation becomes more complicated when other theories of
gravity \cite{bher}, which generalize Einstein's theory, are considered
because the source of the field equations changes and a $\rho +p=0 $ fluid
stress no longer results in a simple de Sitter space-time. The most familiar
example of this sort is the zero-curvature Brans-Dicke universe \cite{BD},
where a $\rho +p=0$ fluid is no longer equivalent to the presence of an
explicit cosmological constant term in the gravitational Lagrangian, and as
such does not produce an asymptotic approach to the de Sitter metric \cite%
{math, la, Bmaeda}. Instead, power-law expansion of the isotropic expansion
scale factor results, with the power becoming infinitely large as the theory
approaches general relativity. This difference arises because, even when $%
\rho $ is constant, the $G\rho $ term in the Friedmann equation falls in
time, as $t^{-2}$, due to the fact that the gravitational `constant' varies
as $G\propto t^{-2}$ \cite{jb3}. Higher-order gravity theories with
Lagrangian contributions that are $O(1/R)$ also allow late-time acceleration
to occur in a variety of different ways in the asymptotic limit of low
4-curvature, $R\rightarrow 0$.

In this paper, we will first consider the approach to the de Sitter metric
in a generalized scalar-tensor theory \cite{nord}, in order to gain some
understanding of what type of accelerated expansion is possible in the
situation where the simplest form of vacuum stress, with an equation of
state $\rho +p=0,$ dominates the expansion dynamics. In most scalar-tensor
theories, the introduction of a vacuum stress will not result in a de Sitter
spacetime. We shall be specifically interested in those theories which omit
late times solutions in which the space-time approaches the de Sitter
solution. We will classify such scalar-tensor theories of gravity by the
speed of asymptotic approach to this de Sitter limit. In all such theories, $%
G$ asymptotes towards a constant and non-zero value. In the second part of
the paper, we consider the asymptotic evolution of scalar-tensor theories in
which the gravitation `constant' vanishes at late times. A general method
for finding solutions of scalar-tensor cosmological models was given in by
Barrow and Mimoso \cite{bmim, bp, jb1}. It is possible to adapt their
generating-function method to study asymptotics, as was done for another
problem by Deruelle et al \cite{der}, but we shall adopt a more direct
approach. We classify the variation of $G(t)$ into three classes and then
use our results to show how a solution of the `Boltzmann brain' problem can
be obtained in a wide range of scalar-tensor cosmologies, which includes the
Brans-Dicke theory as a particular case.

In Section 2, we specify the general field equations for scalar-tensor
theories defined by an arbitrary coupling function $\omega (\phi )$ of the
scalar field. In Section 3, we classify the possible rates of approach to de
Sitter spacetime for different behaviours of $\omega (\phi )$ and provide
the conditions under which a generalized scalar tensor theory will omit a de
Sitter limit. In Section 4 we consider the asymptotic forms for the possible
evolutions of $G(t)$ in scalar-tensor cosmologies when a vacuum stress is
present, using the characteristic $G\propto t^{-2}$ behaviour of Brans-Dicke
theory as a benchmark. We show how theories in which $\phi \rightarrow \infty
$ asymptotically, and the coupling function of the theory, $\omega (\phi )$,
asymptotes to a sufficiently large value, or $\rightarrow \infty$ more
slowly than $\phi^{1/2}$, will not suffer from the `Boltzmann brain' problem
that besets late-time cosmological evolution in general relativistic
cosmologies. In Section 5, we summarize and discuss our results.

\section{Field Equations}

Consider a generalized scalar-tensor theory of gravity described by the
action: 
\begin{equation*}
S=\frac{1}{16\pi }\int d^{4}x\sqrt{-g}\left( \phi R-\omega (\phi )\frac{%
\partial _{a}\phi \partial ^{b}\phi }{\phi }\right) +S_{m}(g_{ab},\psi _{m}),
\end{equation*}%
where we have set $c=\hbar =1$; $S_{m}$ is the matter action, $\psi _{m}$
labels the matter fields. $g_{ab}$ is the metric, $\phi $ is a scalar field
and $\omega (\phi )$ is an arbitrary function that must be specified to fix
the gravity theory. We define the energy-momentum tensor of matter to be $%
T_{ab}$. By varying the action with respect to the metric, and then with
respect to $\phi $, we obtain the field equations and conservation
equations: 
\begin{eqnarray}
&G_{ab}=\frac{8\pi }{\phi }T_{ab}+\frac{\omega }{\phi ^{2}}\left( \partial
_{a}\phi \partial _{b}\phi -\frac{1}{2}g_{ab}(\partial \phi )^{2}\right) +%
\frac{1}{\phi }\left( \nabla _{a}\nabla _{b}\phi -\square \phi \right) ,& \\
&\square \phi +\frac{\omega ^{\prime }}{3+2\omega }\left( \nabla \phi
\right) ^{2}=\frac{8\pi T}{3+2\omega },& \\
&T^{a}{}_{b;a}=0.&
\end{eqnarray}%
The familiar Brans-Dicke theory arises when the arbitrary coupling function $%
\omega (\phi )$ is taken to be a constant. Under the assumptions of
homogeneity and isotropy and spatial flatness the metric becomes: 
\begin{equation*}
\mathrm{d}s^{2}=\mathrm{d}t^{2}-a^{2}(t)\left[\frac{\mathrm{d} r^2}{1-kr^2}
+ r^2 \mathrm{d} \Omega^2\right].
\end{equation*}%
Here $k$ determines the curvature of the $t=\mathrm{const}$ slices. We
assume that the universe is filled with some vacuum 'dark energy' with
density $\rho _{0}$ and equation of state $\rho _{0}=-p_{0}$, and that the
rest of the matter has energy density $\rho _{1}=\rho _{0}K(y)$ and pressure 
$p_{1},$ where $y=\ln a$ and $\lim_{y\rightarrow \infty
}K(y)=\lim_{y\rightarrow \infty }K_{y}(y)=0$. The conservation equations
require that $\rho _{0}=\mathrm{const}$ and that $p_{1}/\rho _{0}=-\left(
K^{\prime }/3+K\right) $. The gravitational field equations then reduce to: 
\begin{eqnarray}
\frac{3\dot{a}^{2}}{a^{2}} &=&\frac{8\pi \rho _{0}}{\phi }\left( 1+K\right) +%
\frac{\omega (\phi )}{2}\frac{\dot{\phi}^{2}}{\phi ^{2}}-\frac{3\dot{a}}{a}%
\frac{\dot{\phi}}{\phi }-\frac{3k}{a^{2}}, \\
\frac{2\ddot{a}}{a}+\frac{\dot{a}^{2}}{a^{2}} &=&\frac{8\pi \rho _{0}}{\phi }%
\left( 1+K_{y}/3+K\right) -\frac{\omega (\phi )}{2}\frac{\dot{\phi}^{2}}{%
\phi ^{2}}-\frac{\ddot{\phi}}{\phi }-\frac{2\dot{a}}{a}\frac{\dot{\phi}}{%
\phi }-\frac{k}{a^{2}}.
\end{eqnarray}

\section{The De Sitter Limit}

\begin{proposition}
In order for there to be an approach to de Sitter at large 4-volumes, where $%
\phi \rightarrow \phi _{\infty }$, the coupling function $\omega (\phi )$
must diverge faster than $|\phi _{\infty }-\phi |^{-1+\epsilon }$ for all $%
\epsilon >0$ as $\phi \rightarrow \phi _{\infty }\neq 0$; $\phi_{\infty} <
\infty$. \label{prop1}
\end{proposition}

We define $\dot{a}/{a}\equiv H=H_{\infty }/(1-F(y))$, $\phi _{\infty }=8\pi
\rho _{0}/3H_{\infty }^{2}$, $k=\gamma H_{\infty }^{2}$, and $\varphi =(\phi
_{\infty }-\phi )/\phi _{\infty }$. The essential Einstein equations can
then be rearranged to read: 
\begin{eqnarray}
\frac{2F_{y}}{1-F}\left[ 1-\varphi -\frac{1}{2}\varphi _{y}\right]
&=&\varphi _{yy}-\varphi _{y}-\frac{\omega \varphi _{y}^{2}}{(1-\varphi )}
\label{eqn1} \\
&&+(1-F)^{2}\left( K_{y}+2\gamma (1-\varphi )e^{-2y}\right) ,  \notag \\
\frac{\omega \varphi _{y}^{2}}{6(1-\varphi )} &=&2F-F^{2}-\varphi -\varphi
_{y}  \label{eqn2} \\
&&-(1-F)^{2}\left( K-\gamma (1-\varphi )e^{-2y}\right) ,  \notag
\end{eqnarray}%
These equations are equivalent to the pair: 
\begin{eqnarray}
\frac{2F_{y}}{1-F}\left[ 1-\varphi -\frac{1}{2}\varphi _{y}\right]
+6(2F-F^{2}) &=&\varphi _{yy}+5\varphi _{y}+6\varphi  \label{eqn3} \\
&&+(1-F)^{2}\left( K_{y}+6K-4\gamma (1-\varphi )e^{-2y}\right)  \notag \\
-\frac{\omega _{\varphi }\varphi _{y}^{2}}{3+2\omega } &=&\frac{\left(
12+12K+3K_{y}\right) (1-F)^{2}}{3+2\omega }  \label{eqn4} \\
&&+\varphi _{yy}+\varphi _{y}\left( 3+\frac{F_{y}}{1-F}\right) .  \notag
\end{eqnarray}

For there to be a de Sitter limit we need $\lim_{y\rightarrow \infty
}H\rightarrow H_{\infty }=\mathrm{const}$ which implies that $%
\lim_{y\rightarrow \infty }F=0$ and also $\lim_{y\rightarrow \infty }F_{y}=0$%
. It is clear from Eq. (\ref{eqn3}) that $F\rightarrow 0$, $F_{y}\rightarrow
0$ implies that $\varphi _{yy}+5\varphi _{y}+6\varphi \rightarrow 0$ because 
$\lim_{y\rightarrow \infty }K(y)=\lim_{y\rightarrow \infty }K_{y}(y)=0,$ and
so $\varphi _{yy},\,\varphi _{y}\,,\varphi \rightarrow 0$. In order to prove
the proposition \ref{prop1}, we solve Eqs. (\ref{eqn1}-\ref{eqn4})
asymptotically in the limit $y\rightarrow \infty $ for general $F(y)$. We
then find the asymptotic form that $\omega (\phi )$ must take for de Sitter
spacetime to emerge in this limit.

We define: 
\begin{equation*}
M(y)=\frac{1}{(1-F)^{2}}-1-\left( K-\gamma e^{-2y}\right) .
\end{equation*}%
We then have: 
\begin{equation*}
\frac{2F_{y}}{1-F}+6(2F-F^{2})-(1-F)^{2}\left( K_{y}+6K-4\gamma
e^{-2y}\right) =(1-F)^{2}(N_{y}+6N),
\end{equation*}%
and so Eqs. (\ref{eqn1}) and (\ref{eqn3}) become: 
\begin{eqnarray}
(1-F)^{2}M_{y}(y) &=&\varphi _{yy}-\varphi _{y}\left[ 1-\frac{F_{y}}{1-F}%
\right] -\frac{\omega \varphi _{y}^{2}}{1-\varphi }  \label{eqn5} \\
&&+\varphi \left[ \frac{2F_{y}}{1-F}-2\gamma (1-F)^{2}e^{-2y}\right] , 
\notag \\
(1-F)^{2}\left( M_{y}+6M\right) &=&\varphi _{yy}+\left( 5+\frac{F_{y}}{1-F}%
\right) \varphi _{y}  \label{eqn6} \\
&&+\left( 6+4\gamma (1-F)^{2}e^{-2y}+\frac{2F_{y}}{1-F}\right) \varphi . 
\notag
\end{eqnarray}%
We consider three cases: $\lim_{y\rightarrow \infty }M_{y}/M=0$, $%
\lim_{y\rightarrow \infty }M_{y}/M=-q$ for some $0<q<\infty $ and $%
\lim_{y\rightarrow \infty }M_{y}/M\rightarrow 0$. If $K=\gamma =0$ then
these cases represent, respectively: slower than power-law approach (of $%
H-H_{\infty }$) to de Sitter, power-law approach to de Sitter, and faster
than power-law approach to de Sitter.

\subsection{Slow approach: $M_{y}/M\rightarrow 0$}

We begin by considering those cases in which $M_{y}/M\rightarrow 0$ as $%
y\rightarrow \infty $. If $K=\gamma =0$ this would imply $F_{y}/F\rightarrow
0$ and so $|H-H_{\infty }|\rightarrow 0$ more slowly than $a^{-r}$ for any $%
r>0$ i.e. $a^{r}\left\vert H-H_{\infty }\right\vert \rightarrow \infty $.
Asymptotically, Eq. (\ref{eqn6}) reduces to; 
\begin{equation*}
6M(y)\sim \varphi _{pp}+5\varphi _{p}+6\varphi ,
\end{equation*}%
and so 
\begin{equation}
\varphi \sim M(y).
\end{equation}%
Asymptotically, Eq. (\ref{eqn5}) then gives: 
\begin{equation}
-\omega _{\varphi }\varphi _{y}\sim 2-2\frac{\varphi F_{y}}{\varphi _{y}}.
\label{eqn7}
\end{equation}%
Now if $\varphi _{y}/\varphi F_{y}\rightarrow A=\mathrm{const}\neq 0$ as $%
y\rightarrow \infty $ then $\ln \varphi \sim AF+\mathrm{const}$, and we can
clearly see that we cannot have both $F\rightarrow 0$ and $\varphi
\rightarrow 0$ as $y\rightarrow 0$ as we have required for any finite $A$.
Asymptotically, (\ref{eqn4}) gives: 
\begin{equation}
-\omega _{\varphi }\varphi ^{2}-12\sim 6\omega \varphi _{y}.  \label{eqn8}
\end{equation}%
If $\lim_{y\rightarrow 0}1/(\omega \varphi _{y})=0,$ then 
\begin{equation*}
-\frac{\omega _{\varphi }\varphi _{y}}{\omega }\sim 6,
\end{equation*}%
and so $\omega \sim Ae^{-6y}$ for some $A,$ and hence 
\begin{equation*}
\lim_{y\rightarrow \infty}\frac{1}{\omega \varphi _{y}}=\lim_{y\rightarrow
\infty }\frac{e^{6y}}{\varphi }\frac{\varphi }{\varphi _{y}}\rightarrow \pm
\infty
\end{equation*}%
by $\lim_{y\rightarrow \infty}\varphi _{y}/\varphi =0$. It follows then that
we cannot have $\lim_{y\rightarrow \infty}1/(\omega \varphi _{y})=0$ and
hence, by Eq. (\ref{eqn7}), we cannot have $\lim_{y\rightarrow \infty
}\varphi _{y}F_{y}/\varphi =0$.

We must therefore have $\lim_{y \rightarrow \infty} \varphi /(\varphi_{y}
F_{y} )= 0$, and so by both Eqs. (\ref{eqn7}) \& (\ref{eqn8}): 
\begin{equation}
\omega \sim -\frac{2}{\varphi_{y}}.
\end{equation}
It follows from $\lim_{y \rightarrow \infty} \varphi_{y}/\varphi = \lim_{y
\rightarrow \infty} M_{y}/M = 0$ that $\lim_{y \rightarrow \infty} \vert
\varphi \omega \vert \rightarrow \infty$. In order to avoid ghosts we must
have $\omega > -3/2$ and so, at late times, we must have $\varphi_{y} < 0$
so that $\omega \rightarrow +\infty$. It follows that $\varphi \rightarrow
0^{+}$. In this case the gravitational constant, $G = 1/ \phi$, behaves as: 
\begin{equation}
G(t) \sim G_{\infty}(1+ \varphi),
\end{equation}
where $G_{\infty} = \phi_{\infty}^{-1}$. It follows that $G(t) > G_{\infty}$
at finite time and that the limiting value of $G$ is approached from above.

\subsection{Power-law approach: $\lim_{y\rightarrow \infty }M_{y}/M=-q$, $%
0<q<\infty $}

We now consider those cases where $M_{y}/M\rightarrow -q<0$ as $y\rightarrow
\infty $. This implies that $M(y)=f(y)e^{-qy}$ where $f_{y}/f\rightarrow 0$
as $y\rightarrow \infty $. If $K=\gamma =0$ then this equates to $%
a^{r}\left\vert H-H_{\infty }\right\vert \rightarrow 0$ for all $r>q\neq
\infty $ and $a^{r}\left\vert H-H_{\infty }\right\vert \rightarrow 0$ for
all $r<q\neq 0$. In these cases, Eq. (\ref{eqn6}) gives: 
\begin{equation*}
\left( (6-q)f(y)+f_{y}\right) e^{-qy}\sim \varphi _{yy}+5\varphi
_{y}+6\varphi ,
\end{equation*}%
which, if $q\neq 6$, has the solution: 
\begin{equation*}
\varphi \sim c_{0}e^{-2y}+c_{1}e^{-3y}+\frac{(6-q)fe^{-qy}}{(3-q)(2-q)}.
\end{equation*}%
If $\left( c_{0}=c_{1}=0\right) $, $\left( 0<q<2\right) $ and/or $(c_{0}=0$
and $2<q<3)$ then from Eq. (\ref{eqn5}) we have 
\begin{equation}
-qf(y)e^{-qy}-\left( q+q^{2}\right) \varphi =-q^{2}\omega \varphi ^{2}.
\end{equation}%
It follows that: 
\begin{equation}
\omega \sim \frac{12}{q(6-q)\varphi }.  \label{omegagen}
\end{equation}%
If $c_{0}\neq 0$ and $q>2$ then: 
\begin{equation*}
\varphi \sim c_{0}e^{-2y},
\end{equation*}%
and from Eq. (\ref{eqn5}) we have: 
\begin{equation}
\omega \sim \frac{3}{2\varphi }.  \label{omega2}
\end{equation}%
If $c_{0}=0$, $c_{1}\neq 0$ and $q>3$ then: 
\begin{equation*}
\varphi \sim c_{1}e^{-3y},
\end{equation*}%
and Eq. (\ref{eqn5}) gives: 
\begin{equation}
\omega \sim \frac{4}{3\varphi }.  \label{omega3}
\end{equation}%
If $q=2$ then: 
\begin{equation*}
\varphi \sim 4yM(y),
\end{equation*}%
and Eq. (\ref{eqn5}) gives Eq. (\ref{omega2}). If $q=3$ then: 
\begin{equation*}
\varphi \sim -3yM(y),
\end{equation*}%
and once again $\omega $ is described by Eq. (\ref{omega3}).

If $q\neq 6$, we have found that the behaviour of $\omega $ is given either
by Eq. (\ref{omegagen}), Eq. (\ref{omega2}) or Eq. (\ref{omega3}). In all
three sub-cases $\varphi \omega \rightarrow \mathrm{const}$ as $y\rightarrow
\infty $, which is consistent with proposition \ref{prop1}. Additionally, we
see that if we define $\lim_{r\rightarrow \infty }\varphi _{y}/\varphi =-r,$
then if $0<r<6$ we have $\varphi \rightarrow 0^{+}$ and so as $t\rightarrow
\infty $, $G(t)$ tends to its limiting value from above. If $r>6$ then $%
\varphi \rightarrow 0^{-}$ and $G(t)$ tends to its limiting value from below.

We must still deal with the $q=6$ case. If $q=6$ then by Eq. (\ref{eqn6}): 
\begin{equation*}
f_{y}e^{-6y}\sim \varphi _{yy}+5\varphi _{y}+6\varphi ,
\end{equation*}%
where $f_{y}/f\rightarrow 0$. Now we therefore have $\varphi \sim
c_{0}e^{-2y}+c_{1}e^{-3y}+g(y)e^{-6y}$ where $g$ solves: 
\begin{equation}
f_{y}\sim g_{yy}-7g_{y}+12g.  \label{eqnfg}
\end{equation}%
such that if $f_{y}=0$ then $g=0$. If $c_{0}\neq 0$ or $(c_{0}=0$ and $%
c_{1}\neq 0)$ then we revert to one of the cases considered above and $%
\omega $ is given, respectively, by Eq. (\ref{omega2}) or Eq. (\ref{omega3}%
). We therefore take $c_{0}=c_{1}=0$. It follows then from the requirement
that $\lim_{\rightarrow \infty }f_{y}/f=0$ that 
\begin{equation*}
\lim_{y\rightarrow \infty }(g_{yy}/f)=\lim_{y\rightarrow \infty
}(g_{y}/f)=\lim_{y\rightarrow \infty }(g/f)=0.
\end{equation*}%
Thus, we have 
\begin{equation*}
\lim_{y\rightarrow \infty }\varphi _{yy}/M_{y}=\lim_{y\rightarrow \infty
}\varphi _{y}/M_{y}=\lim_{y\rightarrow \infty }\varphi /M_{y}=0.
\end{equation*}%
Eq. (\ref{eqn5}) therefore gives: 
\begin{equation}
\omega \sim \frac{6f(y)e^{-6y}}{\varphi _{y}^{2}}=\frac{6}{\varphi }\left[ 
\frac{f(y)g(y)}{(g_{y}-6g)^{2}}\right] .
\end{equation}%
So. if $g_{y}/g\rightarrow 0$, we have 
\begin{equation*}
|\varphi \omega |\rightarrow \frac{1}{6}\left\vert \frac{f}{g}\right\vert
\rightarrow \infty ,
\end{equation*}%
and so $\omega $ diverges faster than $1/\varphi $. If $g_{y}/g\rightarrow
-s,$ with $s<0,$ we cannot have $f_{y}/f\rightarrow 0$ as required, and so $%
s>0$. Thus, 
\begin{equation*}
|\varphi \omega |\rightarrow \frac{6}{(6+s)^{2}}\left\vert \frac{f}{g}%
\right\vert \rightarrow \infty ,
\end{equation*}%
and once again $\omega $ diverges faster than $1/\varphi $. Finally, if $%
g_{y}/g\rightarrow -\infty ,$ then we write $g=e^{-b(y)}$ where $%
b,\,b_{y}\rightarrow \infty $. Now $\lim_{y\rightarrow \infty }\left(
yb_{y}\right) ^{-1}=-\lim_{y\rightarrow \infty }b_{yy}/b_{y}^{2}=0$, and so $%
\lim_{y\rightarrow \infty }g_{yy}g/g_{y}^{2}=1$. By Eq. (\ref{eqnfg}) we
also have $g_{yy}\sim f_{y}$ and so: 
\begin{equation*}
\left\vert \varphi \omega \right\vert \sim \left\vert \frac{6fg}{g_{y}^{2}}%
\right\vert \sim \frac{6f}{g_{yy}}\sim \frac{6f}{f_{y}}\rightarrow \infty
,\qquad \mathrm{as}\,y\rightarrow \infty .
\end{equation*}%
It follows that if $q=6$ then $\omega $ diverges faster than $1/\varphi $.

We have shown that if $M_{y}/M\rightarrow -q$ then $\omega $ either diverges
as $1/\varphi $ or, if $q=6$, it is possible for $\omega $ to diverge faster
than $1/\varphi $. In all cases then, we have, as proposed \ref{prop1}, that 
$\omega $ diverges faster than $|\varphi |^{-1+\epsilon }$ for all $\epsilon
>0$.

\subsection{Fast approach: $M_{y}/M\rightarrow -\infty $}

We now consider those cases where $M_{y}/M\rightarrow -\infty $ as $%
y\rightarrow \infty $. It follows from Eq. (\ref{eqn6}) that $\varphi \sim
c_{0}e^{-2y}+c_{1}e^{-3y}+\bar{\varphi}$ where 
\begin{equation}
M_{y}\sim \bar{\varphi}_{yy}\Rightarrow \bar{\varphi}_{y}\sim M.
\end{equation}%
If $c_{0}\neq 0,$ then Eq. (\ref{eqn5}) tells us that the behaviour of $%
\omega $ is described by Eq. (\ref{omega2}). If $c_{0}=0$ and $c_{1}\neq 0$
then the behaviour of $\omega $ is given by Eq. (\ref{omega3}). In both
cases $\omega $ diverges as $\varphi ^{-1}$. We now consider the case $%
c_{0}=c_{1}=0$ so $\varphi =\bar{\varphi}$. From Eq. (\ref{eqn4}), we have 
\begin{equation}
-\omega _{\varphi }\varphi _{y}^{2}\sim 12+2\omega \varphi _{yy}.
\end{equation}%
We define $J$ by $J_{y}=M$ and $J\rightarrow 0$ as $y\rightarrow \infty $,
and define $J=e^{-g(y)}$. It follows from $J_{y}/J\rightarrow -\infty $ and $%
J\rightarrow 0$ that $g,\,g_{y}\rightarrow \infty $. Now $\lim_{y\rightarrow
\infty }\left( yg_{y}\right) ^{-1}=-\lim_{y\rightarrow \infty
}g_{yy}/g_{y}^{2}=0$ and so: 
\begin{equation}
\frac{J_{yy}J}{J_{y}^{2}}=1+\frac{g_{yy}}{g_{y}}^{2}\rightarrow 1.
\end{equation}%
Since $\varphi _{y}\sim M$ we have $\varphi \sim J$ and so, using the above
relation, we have 
\begin{equation}
\left( 2\omega +\omega _{\varphi }\varphi \right) \sim -\frac{12\varphi }{%
\varphi _{y}^{2}}.
\end{equation}%
We now defined $\epsilon _{0}$ by $|\varphi ^{1-\epsilon }\omega
|\rightarrow 0$ for all $\epsilon <\epsilon _{0}$ and $\rightarrow \infty $
for all $\epsilon >\epsilon _{0}$. We then have: 
\begin{equation*}
\omega \sim -\frac{12\varphi }{(1+\epsilon _{0})\varphi _{y}^{2}}.
\end{equation*}%
Since $\varphi _{y}/\varphi \rightarrow -\infty $, the above expression
gives us $\lim_{y\rightarrow \infty }\varphi \omega =0$. From the definition
of $\epsilon _{0}$ then, we must have $\epsilon _{0}\geq 0$. Now clearly,
for all $\epsilon >0,$ we have 
\begin{equation*}
\lim_{y\rightarrow \infty }\frac{\varphi ^{\epsilon /2}}{y}=0,
\end{equation*}%
but then 
\begin{equation*}
\lim_{y\rightarrow \infty }\frac{\varphi ^{\epsilon /2}}{y}=\frac{\epsilon }{%
2}\lim_{y\rightarrow \infty }\frac{\varphi _{y}}{\varphi ^{1-\epsilon }}=0.
\end{equation*}%
Thus, for all $\epsilon >0$ as $y\rightarrow \infty $, we have: 
\begin{equation*}
|\varphi |^{1-\epsilon }|\omega |=\left( \frac{\varphi _{y}}{\varphi
^{1-\epsilon /2}}\right) ^{-2}\rightarrow \infty .
\end{equation*}%
Therefore, by the definition of $\epsilon _{0}$ we must have $\epsilon
_{0}\leq 0$, but since we also found that $\epsilon _{0}\geq 0$, and it then
follows that $\epsilon _{0}=0$.

In this fast-approach case then, we have: 
\begin{equation}
\omega \sim -\frac{12\varphi }{\varphi _{y}^{2}}.
\end{equation}%
Since $\varphi _{y}/\varphi \rightarrow -\infty $, we have that $\omega $
diverges faster than $1/\varphi $ in that limit, but since we also found
that $\epsilon _{0}=0$, we have that $|\varphi ^{1-\epsilon }\omega
|\rightarrow \infty $ for all $\epsilon >0$, and so $\omega $ diverges
faster than $|\varphi |^{-1+\epsilon }$ for all $\epsilon >0$.

We also note that for $\omega \rightarrow +\infty $, and hence to avoid
ghosts, we must have $\varphi \rightarrow 0^{-}$ as $y\rightarrow \infty $.
The effective gravitational `constant' $G(t)\sim G_{\infty }(1+\varphi )$
therefore approaches its limiting value from below in the late-time limit.

We have shown that if $M_{y}/M \rightarrow -\infty$ then, as per proposition %
\ref{prop1}, $\omega$ diverges faster than $\vert \varphi
\vert^{-1+\epsilon} $ for all $\epsilon > 0$.

\subsection{Summary}

We have shown that proposition \ref{prop1} holds in each of the three
possible cases. We have therefore proved proposition \ref{prop1}. For there
to be a de Sitter asymptote we therefore require that $\omega $ diverges
faster than $|\phi _{\infty }-\phi |^{-1+\epsilon }$ for all $\epsilon >0$
as $y\rightarrow \infty $, $\phi \rightarrow \phi _{\infty }\neq 0$. Note
that this is a more stringent condition than one might expect from
considering linear scalar-tensor corrections to the metric around a
spherically symmetric body. In the PPN formalism, scalar-tensor corrections
to general relativity vanish as $\omega \rightarrow \infty $ and $\omega
^{\prime }/\omega ^{3}\rightarrow 0$. The latter condition might lead one to
expect a de Sitter cosmological limit if $\omega $ diverges faster than $1/%
\sqrt{|\phi _{\infty }-\phi |}$, however we have found that this is not the
case. In summary: for a given theory of gravity, specified by $\omega (\phi
),$ a de-Sitter limit for late-time Friedmann cosmology requires that there
exists some $\phi _{\infty }\in (0,\infty )$ such that as $\phi \rightarrow
\phi _{\infty }$ 
\begin{equation*}
\omega \rightarrow \infty ,\qquad \frac{\omega ^{\prime }}{\omega
^{2+\epsilon }}\rightarrow 0,
\end{equation*}%
for all $\epsilon >0$. If we define $-r=\lim_{y\rightarrow \infty }\varphi
_{y}/\varphi $ then $G(t)\rightarrow G_{\infty }^{+}$ for $r<6$ and $%
G(t)\rightarrow G_{\infty }^{-}$ for $r>6$.

\section{Decaying Gravity}

In theories with asymptotically de Sitter behaviour, $G(t)\rightarrow 
\mathrm{const}\neq 0$ in that limit. Another interesting limit to consider
is that in which $G(t)\rightarrow 0$ at late times. In particular, when the
equation of state is $\rho =-p$ an exact zero-curvature ($k=0$) Friedmann
solution of Brans-Dicke theory with this property is known. In this
solution, $\omega =\mathrm{const}$, $a\propto t^{\omega +1/2}$ and $%
G(t)\propto t^{-2}$. If $\omega \gtrsim 40,000$ today, as implied by
tracking data for the Cassini spacecraft \cite{bounds}, then the possibility
that the evolution of our universe is described by such a solution is not
ruled out by observational or experimental evidence. Although the $G(t)$
evolution is strong, and appears to be $\omega $ independent, it is better
to examine the $G(a)$ $\propto a^{-4/(2\omega +1)}$ evolution, where the $%
\omega $ dependence appears explicitly and the $\omega \rightarrow \infty $
general relativity limit is evident. In order to classify the late-time
behaviour of all such solutions, we define $G=G_{0}\chi (y)$, $H=H_{0}F(y)$, 
$k/a^{2}=\gamma H_{0}^{2}e^{-2y}$ where $8\pi G_{0}\rho _{0}/3H_{0}^{2}=1$;
and as above, we have taken $y=\ln a$. With these definitions, the Einstein
equations are now equivalent to: 
\begin{eqnarray}
\left( 1-\frac{\omega }{6}\frac{\chi _{y}^{2}}{\chi ^{2}}-\frac{\chi _{y}}{%
\chi }\right) &=&\frac{\chi }{F^{2}}\left( 1+K\right) -\frac{\gamma }{F^{2}}%
e^{-2y},  \label{sGeqn1} \\
\frac{2F_{y}}{F}\left( 1-\frac{\chi _{y}}{2\chi }\right) +3 &=&\frac{3\chi
\left( 1+K_{y}/3+K\right) }{F^{2}}-\left( \frac{\omega }{2}+2\right) \frac{%
\chi _{y}^{2}}{\chi ^{2}}+\frac{\chi _{yy}}{\chi }+2\frac{\chi _{y}}{\chi }-%
\frac{\gamma e^{-2y}}{F^{2}}.  \label{sGeqn2}
\end{eqnarray}%
We can combine Eqs. (\ref{sGeqn1}) and (\ref{sGeqn2}) to give: 
\begin{eqnarray}
\frac{2F_{y}}{F}\left( 1-\frac{\chi _{y}}{2\chi }\right) +6 &=&\frac{6\chi
\left( 1+K_{y}/6+K\right) }{F^{2}}-\frac{2\chi _{y}^{2}}{\chi ^{2}}+\frac{%
\chi _{yy}}{\chi }+5\frac{\chi _{y}}{\chi }-\frac{4\gamma e^{-2y}}{F^{2}},
\label{sGeqn3} \\
\frac{2F_{y}}{F}\left( 1-\frac{\chi _{y}}{2\chi }\right) &=&\frac{\chi K_{y}%
}{F^{2}}+\frac{2e^{-2y}}{F^{2}}+\frac{\chi _{yy}}{\chi }-\frac{\chi _{y}}{%
\chi }-\left( \omega +2\right) \frac{\chi _{y}^{2}}{\chi ^{2}}.
\label{sGeqn4}
\end{eqnarray}

We now divide the possible rates of change of $G(t)$ towards its asymptotic
value into three classes and study each in turn.

\subsection{Slow Approach: $G_{y}/G\rightarrow 0$}

If $G\rightarrow 0$ more slowly than $a^{-r}$ for any $r>0$, then $\chi
_{y}/\chi \rightarrow 0$ at late times, and Eq. (\ref{sGeqn3}) gives 
\begin{equation}
\chi \sim F^{2}+F_{y}F/3,
\end{equation}%
and so using $\lim_{y\rightarrow 0}\chi _{y}/\chi =0$ we have: 
\begin{equation}
\chi \sim F^{2}.  \label{slowFeqn}
\end{equation}%
We can now use Eq. (\ref{sGeqn4}) to find $\omega (\phi )$: 
\begin{equation*}
\omega \sim -\frac{F}{F_{y}}\left( 1-\frac{K_{y}F}{4F_{y}}\right) \sim -%
\frac{2\chi }{\chi _{y}}\left( 1-\frac{K_{y}\chi }{2\chi _{y}}\right) .
\end{equation*}%
Now, assume that $\chi _{y}/\chi K_{y}\rightarrow \alpha $ for some $\alpha $
as $y\rightarrow \infty $. This would give $\chi \sim B\exp (\alpha K)$ for
some $B\neq 0$. Given that $K\rightarrow 0^{+}$ as $y\rightarrow \infty $
and $\chi \rightarrow 0,$ we cannot have $\alpha >-\infty $. Thus we must
have $K_{y}\chi /\chi _{y}\rightarrow 0$ as $y\rightarrow \infty $, and so
whatever $K$ is: 
\begin{equation}
\omega \sim -\frac{2\chi }{\chi _{y}}.  \label{slowomegaeqn}
\end{equation}%
We note that as $y\rightarrow \infty $, $\omega \rightarrow \infty $, and
also that 
\begin{equation*}
\omega ^{\prime }\sim \frac{G_{0}\omega ^{3}}{4}\left( \chi _{yy}-\frac{\chi
_{y}^{2}}{\chi }\right) ,
\end{equation*}%
and so 
\begin{equation*}
\lim_{y\rightarrow \infty }\frac{\omega ^{\prime }}{\omega ^{3}}=0.
\end{equation*}%
For these solutions: 
\begin{equation*}
G(t)\sim G_{0}F^{2}=\frac{G_{0}H^{2}}{H_{0}^{2}}=\frac{3H^{2}}{8\pi \rho }.
\end{equation*}%
Furthermore, it is clear that any scalar-tensor theory where $\omega
\rightarrow \infty $ and $\omega ^{\prime }/\omega ^{3}\rightarrow 0$ as $%
\phi \rightarrow \infty $ will have such a late-time solution -- since once $%
\omega (\phi )$ is known, the formula $\omega \sim -2\chi /\chi _{y}$ can be
used to find $\chi _{y}$ and hence $H(y)$.

All bounds from local tests of gravity are satisfied in the limit $\omega
\rightarrow \infty $ and $\omega ^{\prime }/\omega ^{3}\rightarrow 0$.
Theories with stable late-time solutions of this class are therefore
physically viable. We note that at late times, 
\begin{equation*}
\Omega _{\Lambda }=\frac{8\pi \rho G}{3H^{2}}\rightarrow 1,
\end{equation*}%
however the spacetime is not not de Sitter since gravitation `constant' is
decaying i.e. $G\rightarrow 0$.

\subsection{Power-law approach: $\lim_{y \rightarrow \infty} G_{y}/G = -q$}

We now consider solutions where $a^{r}G\rightarrow 0$ for $r<q$ and $%
\rightarrow \infty $ for $r>q$. Eq. (\ref{sGeqn3}) gives: 
\begin{equation}
(q+3)(q+2)\sim \frac{6\chi }{F^{2}}-\frac{4\gamma e^{-2y}}{F^{2}}-\frac{%
2F_{y}}{F}(1+q/2).
\end{equation}%
Since we cannot have $q<0$, it follows that we must have $\lim_{y\rightarrow
\infty }F_{y}/F=-r$ for some $r>0$, and $\gamma e^{-2y}/\chi \rightarrow 0$
as $y\rightarrow 0$, which is certainly the case if $0<q<2$ and/or $\gamma
=0 $, then we must have $F^{2}\propto \chi $ in the $y\rightarrow \infty $
limit. Therefore, in this sub-case $2r=q$ and 
\begin{equation}
\chi \sim \frac{(q+6)(q+2)F^{2}}{12}.
\end{equation}%
From Eq. (\ref{sGeqn1}), then we have $\lim_{y\rightarrow \infty }\omega
=\omega _{\infty }$ and 
\begin{equation}
\omega _{\infty }=\frac{2}{q}-\frac{1}{2}>-1/2.
\end{equation}

If $|\gamma e^{-2y}/\chi |\rightarrow \infty $, which certainly requires $%
q\geq 2$ and $\gamma \neq 0$, Eq. (\ref{sGeqn3}) gives: 
\begin{equation*}
(q+3-r)(q+2)\sim \frac{-4\gamma e^{-2y}}{F^{2}}.
\end{equation*}%
Since $q>2$, we must therefore have $r=1$ and also $F^{2}\sim -4\gamma
e^{-2y}/(q+2)^{2}$. For such solutions to exist we must therefore have $%
\gamma <0$. From Eq. (\ref{sGeqn1}) we have: 
\begin{equation*}
1+q-\frac{\omega q^{2}}{6}\sim \frac{(q+2)^{2}}{4},
\end{equation*}%
Thus, we have $\lim_{y\rightarrow \infty }\omega =\omega _{\infty }$ where 
\begin{equation*}
\omega _{\infty }=-\frac{3}{2}.
\end{equation*}

Finally, if $q=2$ and $\chi e^{2y}\rightarrow a_{0}=\mathrm{const}\neq 0$,
then Eq. (\ref{sGeqn3}) gives $r=1$ and 
\begin{equation}
\frac{3}{2}\chi \sim \gamma e^{-2y}+4F^{2}.
\end{equation}%
Hence, we must have $3a_{0}/2>\gamma $. We define $\gamma =b_{0}a_{0},$ and
then from Eq. (\ref{sGeqn1}) we have $\lim_{y\rightarrow \infty }\omega
=\omega _{\infty },$ where 
\begin{equation}
\omega _{\infty }=\frac{3(1+2b_{0})}{2(3-2b_{0})}>-\frac{3}{2}.
\end{equation}%
Observationally acceptable solutions may require that $\omega >40,000$
today. For $\omega _{\infty }>40000$ we would need $q<5\times 10^{-5}$.

\subsection{Fast approach: $|G_{y}/G|\rightarrow \infty $}

The final class of solutions that we consider have $G\rightarrow 0$ more
quickly than $a^{-r},$ for any $r>0$. This implies that $|\chi _{y}/\chi
|\rightarrow \infty $. For such solutions we write $\chi =e^{-g(y)}$ and we
require that $g,\,g_{y}\rightarrow \infty $. This further implies that $\chi
_{yy}\sim \chi _{y}^{2}/\chi $. We also write $F=\exp (-f(y))$. If $\gamma
=0 $ then Eq. (\ref{sGeqn3}) gives 
\begin{equation}
\left( f_{y}-g_{y}\right) g_{y}\sim -6\exp (2f-g).  \label{fasteq}
\end{equation}%
We define $J$, by $J_{y}=F$ and $J\rightarrow 0$ as $y\rightarrow \infty $.
We note that $J\sim F^{2}/F_{y}$ as $y\rightarrow \infty $ since $%
F_{yy}F/F_{y}^{2}\rightarrow 1$. We now make the ansatz: 
\begin{equation*}
e^{g}\sim \lambda J^{2}.
\end{equation*}%
With this definition of $\lambda $: 
\begin{equation*}
g_{y}^{2}\sim 4\lambda e^{2f-g}
\end{equation*}%
Since $g_{y}=2J_{y}/J$ and $F_{y}=J_{yy},$ we have 
\begin{equation*}
\frac{f_{y}}{g_{y}}=\frac{J_{yy}J}{2J_{y}^{2}}\sim \frac{1}{2}.
\end{equation*}%
Eq. (\ref{fasteq}) then gives 
\begin{equation}
g_{y}^{2}\sim 12e^{2f-g}.
\end{equation}%
It follows that $\lambda =\sqrt{3}$ is required. In this case then: 
\begin{equation*}
\chi \sim \frac{\sqrt{3}F^{2}}{F_{y}}.
\end{equation*}%
Using Eq. (\ref{sGeqn1}) we find: 
\begin{equation*}
(1+\frac{2\omega }{3})\rightarrow 0
\end{equation*}%
and so as $y\rightarrow \infty $, 
\begin{equation}
\omega \rightarrow \omega _{\infty }=-\frac{3}{2}.
\end{equation}

If $\gamma \neq 0,$ Eq. (\ref{sGeqn3}) gives $f\sim y+h$ where: 
\begin{equation}
\left( h_{y}-g_{y}\right) g_{y}\sim 4\gamma \exp (2h)
\end{equation}%
We define $L_{y}=e^{h}$ and so as $y\rightarrow \infty $, $L\rightarrow
\infty $. Solutions to this equation require $g\sim \mu L$. It follows from
this that $h_{y}/g_{y}\rightarrow 0$, and so $\mu =\sqrt{-4\gamma }$. Since
we must have $g\rightarrow \infty $, such solutions only exist is $\gamma <0$%
. In this case then: 
\begin{equation}
\chi =e^{-g}\sim e^{-\mu L}.
\end{equation}%
From Eq. (\ref{sGeqn1}), we find: 
\begin{equation*}
-\frac{\mu ^{2}\omega }{6}\sim -\gamma =\mu ^{2}/4,
\end{equation*}%
and so as $y\rightarrow \infty $: $\omega \rightarrow \omega _{\infty }=-3/2$%
.

\subsection{Suppressing "Boltzmann Brains"}

We have shown that solutions in which the gravitational constant decays at
late times i.e. $G\rightarrow 0$, require that, as $G^{-1}\sim \phi
\rightarrow \infty $ either $\omega \rightarrow \infty $ or $\omega
\rightarrow \omega _{\infty }=\mathrm{const}$. Generally speaking, the
smaller $\lim_{\phi \rightarrow \infty }\omega $ is, the faster $G$ decays
at late times. The fastest decay occurs when $\omega _{\infty }=-3/2$.
Observational data may require that$\omega >40,000,$ today \cite{bounds},
and so solutions with small values of $\omega _{\infty }$ are unlikely to
exist in physically viable theories. Theories with very large values of $%
\omega _{\infty }$ are, however, permitted alternatives to standard general
relativity.

The decay of $G$ in similar scalar-tensor theories would occur over time
scales that are large compared to the Hubble time, and at late finite times
the spacetimes they predict would be observationally similar to de Sitter,
although both $G$ and $H$ would be decaying with time. This is an
interesting scenario. If space-time is asymptotically de Sitter, then $%
H\rightarrow H_{0}\neq 0$ at late times. Since the expansion is
accelerating, there exists an event horizon, and as is the case with a black
hole horizon, the de Sitter horizon has a temperature: $T=H_{0}/2\pi =%
\mathrm{const}\neq 0$. The presence of a non-zero minimum temperature in
space-times that asymptote to de Sitter is important, as it means that even
very rare thermal fluctuations are eventually expected to occur. For
instance, the probability that a certain thermal fluctuation occurs goes as $%
e^{-S/T}$ for some action $S$ which does not depend on $T$. The 4-volume of
space-time grows as $\int a^{3}\mathrm{d}t,$ which in de Sitter spacetime
grows as $e^{3H_{0}t}$. Thus, after a certain time $t$, the number of times
that a particular thermal fluctuation is expected to have occurred is $n\sim
e^{-S/T+3Ht}$. Provided $T\rightarrow \mathrm{const}$, $n$ will eventually
be greater than unity, and $n\rightarrow \infty $ asymptotically. A
particularly extreme and topical example of this is the spontaneous
emergence of \textquotedblleft Boltzmann brains\textquotedblright\ in
general -relativistic space-times that asymptote to de Sitter. Self-aware
\textquotedblleft observers\textquotedblright\ could emerge from the vacuum
as a result of thermal fluctuations \cite{bbrain}. In an eternal,
asymptotically de Sitter, universe such \textquotedblleft
observers\textquotedblright\ should vastly outnumber \textquotedblleft
ordinary\textquotedblright\ carbon-based observers such as ourselves. This
conclusion is changed in scalar-tensor cosmologies. 

In scalar-tensor gravity, we have seen that theories in which either $\omega
\rightarrow \omega _{\infty }>40,000$, or $\omega \rightarrow \infty $ as $%
\phi \rightarrow \infty $, admit asymptotic solutions that would be
observationally indistinguishable from de Sitter today, but in which $%
G,H\rightarrow 0$ at late times. Since $G$ and $H$ would, in such theories,
decay only very slowly over a Hubble time, we still have $T\approx H/2\pi $.
The 3-volume of a spatial slice of constant $t$ goes like $%
a^{3}/H^{3}\propto e^{3y-3\ln F}$, and so the number of thermal fluctuations
that are expected goes like: 
\begin{equation}
n\sim \int^{t}e^{-S/T(t)+3y-3\ln F}\mathrm{d}t\propto \int^{y}\left( e^{-%
\frac{2\pi S}{H_{0}F(y)}+3y-4\ln F}\right) \mathrm{d}y\propto \int^{y}\left(
e^{-\frac{2\pi S}{F(y)}+3y}\right) \mathrm{d}y.  \label{neqn}
\end{equation}%
where, as above, $F=H/H_{0}$ and $y=\ln a$.

We have found that in all theories where $\omega \rightarrow \infty $ or $%
\omega \rightarrow \omega _{\infty }>40,000$ as $\phi \rightarrow \infty $
at late times, $F\rightarrow 0$ slower than $e^{-qy}$, where $q\lesssim
5\times 10^{-5}$. Even with this strong restriction, however, there is still
a class of theories for which $F\rightarrow 0$ faster than $1/y$ as $%
y\rightarrow \infty $. Specifically, from Eqs. (\ref{slowFeqn})-(\ref%
{slowomegaeqn}) we find that this will be the case if, as $\phi \rightarrow
\infty $, $\omega (\phi )$ grows more slowly than $\phi ^{1/2}$; Brans-Dicke
theories with $\omega =\mathrm{const}$ certainly satisfy this constraint. In
these theories, as $y\rightarrow \infty $, the integrand in Eq. (\ref{neqn})
would asymptotically tend to zero faster than $e^{-ry}$ for any $r$. Thus, $%
n\rightarrow n_{\infty }=\mathrm{const}<\infty $ at late times. If $S$ is
small enough, then $n_{\infty }\ll 1$. In these cases, even though the
universe would still be eternal, very rare thermal fluctuations like
\textquotedblleft Boltzmann brains" would not be expected to occur even once.

This is another example (see also \cite{Carlip}), of how, if some or all of
the traditional constants of Nature vary slowly with time, we cannot use
current observations of the universe to make definitive statements about the
expected behaviour of the universe in the far future.

\section{Conclusions}

We have investigated two features of the general behaviour of scalar-tensor
gravity theories. Motivated by the need to understand the possible origins
of de Sitter expansion in the early and late periods of the universe's
history we have investigated how it can arise in general scalar-tensor
gravity theories. We considered Friedmann universes filled with a mixture of
a vacuum stress with equation of state $p=-\rho = -\rho_0 $ and other fluids
which have total energy density $\rho _{1}=\rho _{0}K(y)$ and pressure $%
p_{1},$ where $y=\ln a$ and $\lim_{y\rightarrow \infty
}K(y)=\lim_{y\rightarrow \infty }K_{y}(y)=0,$so the vacuum stress dominates
at late times. In scalar-tensor theories with a coupling function $\omega
(\phi ),$ we find that there is asymptotic approach to de Sitter expansion
at late times, where $\phi \rightarrow \phi _{\infty }$, provided the
coupling function $\omega (\phi )$ diverges faster than $|\phi _{\infty
}-\phi |^{-1+\epsilon }$ for all $\epsilon >0$ as $\phi \rightarrow \phi
_{\infty }\neq 0.$ This means that, for a given theory, specified by $\omega
(\phi ),$ there must exist some $\phi _{\infty }\in (0,\infty )$ such that $%
\omega \rightarrow \infty $ and $\omega ^{\prime }/\omega ^{2+\epsilon
}\rightarrow 0$ as $\phi \rightarrow \phi _{\infty }$ in order for
cosmological solutions of the theory to approach de Sitter expansion at late
times. This differs from the conditions required to establish a general
relativity limit that has vanishing corrections to the weak-field PPN
corrections to general relativity in the solar system: $\omega \rightarrow
\infty $ and $\omega ^{\prime }/\omega ^{3}\rightarrow 0.$

Brans-Dicke theory ($\omega =\mathrm{const}$) does not emit a de Sitter
limit in the presence of $p=-\rho $ stress. There is instead power-law
inflation and $G\propto t^{-2}$. With this behaviour in mind we analysed the
possible late time evolution of $G(t)$ in the Friedmann cosmological models
of scalar-tensor theories defined by an arbitrary $\omega (\phi )$ and
divided them into three classes depending upon the rate of decay of $G$ with
the expansion scale factor. The scenarios in which $G$ decays over quickly
(i.e. over a Hubble time or faster) would be difficult to realize in a
manner that was compatible with solar system tests of gravity i.e. $\omega
>40,000$ today. The subset of theories with a physically viable slow decay
were found to be particularly interesting because if, as is the case in
Brans-Dicke theory, $\omega \sim o(\phi ^{1/2})$ as $\phi \rightarrow \infty 
$, then the expected number of extremely rare thermal fluctuations, $n(t)$,
that occur within the visible universe, after a time $t$, would asymptote to
a constant value. In general relativistic de Sitter space-time, $%
n(t)\rightarrow \infty $, which has led some to postulate that a typical
`observer' of our Universe would most likely have arisen out of the vacuum
as a thermal fluctuation. The intrinsic probability of such a "Boltzmann
brain" fluctuation is tiny, but since $n(t)\rightarrow \infty $ in an
eternal de Sitter universe, the number of such observers would grow without
bound. In scalar-tensor theories, like Brans-Dicke, we have shown that the
slow decay of $G(t)$ can prevent this strange situation from occurring with
any significant probability.

\textbf{Acknowledgements}: D.J. Shaw acknowledges support by STFC at
Cambridge University and Queen Mary University.

\end{document}